\documentclass[aip,jcp,amsmath,amssymb,reprint,nofootinbib]{revtex4-1}

\usepackage{graphicx}
\usepackage{bm}
\usepackage[all,2cell]{xy}
\usepackage{chemarrow}
\usepackage{amssymb,amsfonts,amsmath}
\usepackage{color}
\usepackage{siunitx}

\newcommand{\be}{\begin{equation}}
\newcommand{\ee}{\end{equation}}
\newcommand{\bea}{\begin{eqnarray}}
\newcommand{\ba}{\begin{array}}
\newcommand{\eea}{\end{eqnarray}}
\newcommand{\bes}{\begin{subequations}\bea}
\newcommand{\ees}{\eea\end{subequations}}
\newcommand{\ea}{\end{array}}
\newcommand{\bs}[1] {\boldsymbol{#1}}
\newcommand{\tc}[1]{\textcolor{black}{#1}}
\newcommand{\aw}[1]{\textcolor{black}{#1}}

\begin{document}

\title{Dissipation in noisy chemical networks: The role of deficiency}

\author{M. Polettini}
\email{matteo.polettini@uni.lu}
\affiliation{
Complex Systems and Statistical Mechanics, Physics and Materials Science Research Unit, University of Luxembourg, 162a avenue de la Fa\"iencerie, L-1511 Luxembourg (G. D. Luxembourg)}  

\author{A. Wachtel}
\email{artur.wachtel@uni.lu}
\affiliation{
Complex Systems and Statistical Mechanics, Physics and Materials Science Research Unit, University of Luxembourg, 162a avenue de la Fa\"iencerie, L-1511 Luxembourg (G. D. Luxembourg)} 

\author{M. Esposito}
\email{massimilano.esposito@uni.lu}
\affiliation{
Complex Systems and Statistical Mechanics, Physics and Materials Science Research Unit, University of Luxembourg, 162a avenue de la Fa\"iencerie, L-1511 Luxembourg (G. D. Luxembourg)}

\begin{abstract}
We study the effect of intrinsic noise on the thermodynamic balance of complex chemical networks subtending cellular metabolism and gene regulation. A topological network property called {\it deficiency}, known to determine the possibility of complex behavior such as multistability and oscillations, is shown to also  characterize the entropic balance. In particular, when deficiency is zero the average stochastic dissipation rate equals that of the corresponding deterministic model, where correlations are disregarded. In fact, dissipation can be reduced by the effect of noise, as occurs in a toy model of metabolism that we employ to illustrate our findings.  This phenomenon highlights that there is a close interplay between deficiency and the activation of new dissipative pathways at low molecule numbers.
\end{abstract} 

\maketitle

\section{Introduction}

Today, advanced methods in genomics and metabolomics  allow to reconstruct the chemical networks (CN) describing the metabolism of complex organisms \cite{baart,human}. These reconstructions are graphical repositories of thousands of pathways, metabolites, and their stoichiometry.  Much like heat engines, metabolism operates thermodynamic cycles far from equilibrium that transform \tc{low chemical potential} environmental resources into valuable products, at the expense of \tc{high chemical potential} waste. Unlike the working substance of heat engines \tc{(e.g. steam)}, some metabolites, enzymes and cofactors might reach very low concentrations. At this level intrinsic noise, due to discreteness and randomness of molecular collisions, enters into play \cite{elowitz}. Suppression of noise and control of correlations in the abundance of regulatory molecules is crucial for the correct functioning of metabolic networks \cite{levine,tans,paulsson1}. A stochastic description of dynamics and thermodynamics based on jump processes in molecules' populations is then required.

In this direction, the growing field of Stochastic Thermodynamics created the basis for a complete and consistent characterization of irreversibility in small nonequilibrium systems subject to fluctuations. Dissipation is quantified by the rate at which entropy is produced (EPR) and eventually delivered to the environment \cite{broeck}. The theory has been applied to general CNs \cite{ross,seifert,PolEspo14} such as those involved in  gene regulation \cite{ghosh}, cellular computation \cite{mehta}, copolymerization \cite{andrieuxPNAS}, kinetic proofreading \cite{rao}, chemical switches \cite{gaspard}, {\tc{and signal transduction \cite{qian2}}. On the other hand, there is a growing body of mathematical literature linking a CN's topology to its dynamics, and still bearing no thermodynamic interpretation. In particular, it has been understood that a topological number called  {\it deficiency} subtends the onset of complex behavior, such as bistability and oscillations \cite{feinberg,craciun06,acgw}, which are the mechanisms of chemical switches and clocks \cite{tyson}. When intrinsic noise is important, a crucial result by Anderson, Craciun and Kurtz (ACK) \cite{Ander10} relates the deficiency of the CN to steady statistical properties of the chemical mixture.

In this paper we merge stochastic thermodynamics and deficiency theory, via the ACK theorem. We compare the behavior of an arbitrary CN subject to intrinsic noise and that of the corresponding deterministic model without noise, which follows \tc{deterministic rate equations} where correlations between species are neglected. In the limit of large \tc{particle numbers} the deterministic dynamics describes the mode, i.e. the most typical behavior of the system. The difference between the stochastic and the deterministic EPR in the two cases, here named {\it correlation EPR} (previously known as fluctuating EPR, today ambiguous), is known to vanish at steady states for linear CNs where only input/output and conformational changes of a molecule are allowed, and reaction velocities are linear-affine in the molecules' populations \cite{Mou84}.

The main result in this paper is to extend this observation to nonlinear CNs with null deficiency at steady states, and to linear networks at all times. We rely on the following formula for the steady correlation EPR as the weighted difference between the mean and the mode of the reaction velocity $v$,
\bea 
\textrm{correlation EPR}  =  \left(\textrm{mean}\, v  -\textrm{most probable}\,v \right) G, \label{eq:main}
\eea
where $G$ is the free-enthalpy increase. Hence the correlation EPR might be interpreted as a measure of a system's ``propensity to complexity''.

The plan of the paper is as follows. \tc{In Sec.\,\ref{meta}} we provide a simple definition of deficiency with the aid of a toy model of metabolism. More generally, under the assumption that the law of mass-action holds and that the mixture is well-stirred, we illustrate the dynamics and thermodynamics of CNs, in the stochastic \tc{(\ref{stocEPR})} and in the deterministic \tc{(\ref{detEPR})} settings. We then derive the above formula, and by virtue of the ACK theorem (whose proof we briefly sketch in Appendix \ref{app2}) we draw our main conclusion that the correlation EPR vanishes for networks with zero deficiency. Our toy model will finally serve as a testing ground. \tc{We employ it to illustrate through Figs. \ref{bihisto}, \ref{histo} the predictions of the ACK theorem.} Incidentally, the model displays a non-positive correlation EPR, somewhat contrary to the intuition that ``large variability is likely to [\ldots] increase metabolic burden'' \cite{paulsson1}. We give an explanation of this phenomenon in terms of the topology of the state space where stochastic population dynamics occurs, showing that when deficiency is nonzero, for low molecule numbers certain irreversible closed reaction pathways are switched off.

\section{Setup}

\subsection{Notation}

As customary in CN studies, we employ a rather compressed notation. Letting $\bs{\mathrm{X}}$ be the vector of chemical species, a CN is depicted by a set of stoichiometric equations
\bea
\bs{\nu}_{+\rho} \cdot \bs{\mathrm{X}} \quad \autorightleftharpoons{$k_{+\rho}$}{$k_{-\rho}$} \quad \bs{\nu}_{-\rho} \cdot \bs{\mathrm{X}}
\eea
where vectors $\bs{\nu}_{+\rho}$ and $\bs{\nu}_{-\rho}$ contain, respectively, the numbers of molecules of each species being consumed and produced by reaction $\rho$, and $\bs{a} \cdot \bs{b}$ is the scalar product. The stoichiometric vector is defined as $\bs{\nabla}_{\!\rho} := \bs{\nu}_{-\rho} - \bs{\nu}_{+\rho}$, and it describes the net increase of species' populations. The stoichiometric matrix is the matrix that has the stoichiometric vectors as columns, $\nabla = (\bs{\nabla}_{\!\rho} )_{\rho>0}$. We assume that all reactions are strictly reversible, that is, $k_{\pm \rho} > 0$. In sums $\sum_\rho$, index $\rho$ spans over reactions in both directions, unless otherwise specified. Analytic operations between vectors are performed component-wise and imply the scalar product, e.g. $\bs{a}^{\bs{b}} := \prod_i a_i^{b_i}$, $\bs{a}! := \prod_i a_i!$, $\bs{a}\cdot \ln \bs{b} := \sum_i a_i \ln b_i$. Boltzmann's constant $k_B$ is set to unity.

\subsection{\label{meta}From metabolism to deficiency}

\tc{Roughly speaking,} the deficiency of a CN is the number of ``hidden'' closed pathways, or thermdynamic cycles. Let us make this more precise with a simple model inspired by metabolism. Emphasis is on the cycle structure (see\cite{schaft} for a formal introduction). The model reads
\begin{equation}\label{eq:metamodel}
\begin{aligned}
\emptyset & \stackrel{1}{\longrightarrow} \mathrm{N}  \\
\mathrm{N} + m \mathrm{E} & \stackrel{2}{\longrightarrow} (m+n) \mathrm{E} + \mathrm{W}  \\
n \mathrm{E} + \mathrm{W} & \stackrel{3}{\longrightarrow} \emptyset,
\end{aligned}
\end{equation}
where $\emptyset$ signifies the ``environment'' as a whole. The first reaction introduces nutrients $\mathrm{N}$. The second processes the nutrients with the aid of $m$ tokens of energy $\mathrm{E}$ to produce more tokens of energy and waste $\mathrm{W}$, and the third delivers waste and excess energy to the environment.

When all three reactions in the above network are performed in a pathway, a thermodynamic cycle is completed, restoring all concentrations in the system to their initial value at the expense of irreversibly dissipated free enthalpy (entropy production). Correspondingly, the stoichiometric matrix
\bea
\nabla = \left(\ba{cccc} +1 & -1 &  0 \\  0 & +1 & -1 \\ 0 & +n & -n \ea\right)
\eea
admits $\bs{c}  = (1,1,1)^T$ as a right-null vector, $\nabla \bs{c} = 0$ \cite{PolEspo14}.

The crucial step to understand deficiency is to introduce a symbolic representation of the network in terms of {\it complexes}, which are aggregates of species appearing as either reactants or products in a reaction. In our case, the complexes are $\mathrm{Y}_1 = \emptyset,\mathrm{Y}_2= \mathrm{N},\mathrm{Y}_3= \mathrm{N} + m \mathrm{E}, \mathrm{Y}_4= (m+n) \mathrm{E} + \mathrm{W}, \mathrm{Y}_5= \mathrm{W} + n\mathrm{E}$.  We then obtain a representation of the CN as a graph by drawing each reaction as an edge connecting vertices given by the complexes.

For $m= 0$, we notice that  $\mathrm{Y}_2= \mathrm{Y}_3$ and $\mathrm{Y}_4 = \mathrm{Y}_5$ and that a representation of the above network in terms of complexes is a graph consisting of one cycle:
\bea
\ba{c}\xymatrix{
  \mathrm{Y}_1 \ar@{->}^{1}[rr]   & &  \mathrm{Y}_2 \ar@{->}^{2}[dl]  \\ & \mathrm{Y}_4 \ar@{->}^{3}[ul] }\ea.
\eea
Its topology is fully described by its {\it incidence matrix}
\bea
\partial = \left(\ba{cccc}
-1 & 0 & +1  \\
+1 & -1 & 0  \\ 
0 & +1 & -1
\ea\right)
\eea
which admits one right null vector.

For $m>0$ we obtain the representation
\bea
\ba{c}\xymatrix{
\mathrm{Y}_5 \ar@{->}^{3}[r]   &  \mathrm{Y}_1 \ar@{->}^{1}[r]  &  \mathrm{Y}_2}\ea, \qquad \ba{c} \xymatrix{\mathrm{Y}_3  \ar@{->}^{2}[r]  & \mathrm{Y}_4}\ea, \label{eq:motif}
\eea
with incidence matrix
\bea
\partial = \left(\ba{cccc}
-1 & 0 & +1  \\
+1 & 0 & 0  \\ 
0 & -1 & 0  \\
0 & +1 & 0  \\
0 & 0 & -1  \\
\ea\right)
\eea
This graph has no cycles; in fact its incidence matrix admits no right-null vectors.

The {\it deficiency} $\delta$ of a CN is the number of independent closed reaction pathways that cannot be visualized as independent cycles in the graphical representation in terms of complexes, and thus in some sense are ``hidden''. In our example when   $m= 0$ then  $\delta = 0$, otherwise the system is deficient, $\delta = 1$. Notice that null deficiency occurs when the autocatalytic mechanism of reaction 2 is not present.

The general recipe to calculate the deficiency is: (i) write down the stoichiometric matrix $\nabla$ of the network; (ii) write down the incidence matrix $\partial$ of the graph where the reactions are arrows and complexes of reactants distinct vertices of the graph; (iii) then the deficiency is
\bea
\delta = \dim \mathrm{ker}\, \nabla - \dim \mathrm{ker}\, \partial  \geq 0
\eea
where $\dim \mathrm{ker}$ calculates the dimension of the null space. The deficiency is non-negative. In fact one can write
\bea
\nabla = \frac{\partial Y}{\partial X} \partial \label{eq:ker}
\eea
where the entry $(\partial Y/\partial X)_{ij}$ quantifies the amount of species $X_i$ in complex $Y_j$. Since by Eq.\,(\ref{eq:ker}) a right-null vector of $\partial$ is necessarily a right-null vector of $\nabla$, then $\delta \geq 0$.

\subsection{Average stochastic EPR}

\label{stocEPR}

The setup of Markovian population dynamics of chemical species is as follows. The number of molecules in the reactor performs a jump process on the discrete lattice orthant $\mathcal{Z}_{\bs{X}_0}$ of populations that, starting from the initial state $\bs{X}_0$, are reachable by a finite number of reactions\footnote{That is, $\mathcal{Z}_{\bs{X}_0} := \{\bs{X} = \bs{X}_0 + \nabla \bs{n}, \bs{n} \in \mathbb{Z}^R, \bs{X} \geq \bs{0} \}$, sometimes called the {\it stoichiometric compatibility class}, compatible with $\bs{X}_0$.}. According to the law of mass-action, transition $\bs{X} \stackrel{\rho}{\longrightarrow} \bs{X} +  \bs{\nabla}_{\!\rho}$ is performed at rate
\bea
v_\rho(\bs{X}) = k_{\rho} \frac{\bs{X}!}{(\bs{X} - \bs{\nu}_{\rho})!}. \label{eq:LMA}
\eea
The probability  (or {\it ensemble}) $p_t(\bs{X})$ that $\bs{X}$ molecules are present in the reactor at time $t$ obeys the Chemical Master Equation $\dot{p}_t = {L} p_t$ with generator
\begin{multline}
{L}p_t(\bs{X}) = - \sum_{\rho} \Big[ v_{+\rho}(\bs{X}) p_t(\bs{X}) \\ - v_{-\rho}(\bs{X}  +  \bs{\nabla}_{\!\rho}) p_t(\bs{X}+  \bs{\nabla}_{\!\rho}) \Big].  \label{eq:CME}
\end{multline}
Multiplying by, and summing over  $\bs{X}$, one obtains for the mean populations
\bea
\frac{d}{dt}\langle \bs{X} \rangle_t = \sum_{\rho} \bs{\nabla}_{\!\rho} \langle v_\rho(\bs{X}) \rangle_t \label{eq:mean}
\eea
where the average $\langle\,\cdot\,\rangle_t$ is taken with respect to $p_t(\bs{X})$. The equation is not closed, as it involves higher moments on the right-hand side. 

For finite $\mathcal{Z}_{\bs{X}_0}$, it can be proven that any ensemble supported on $\mathcal{Z}_{\bs{X}_0}$ evolves towards a unique steady ensemble $p_\infty$ such that ${L} p_\infty = 0$. We assume that for unbounded $\mathcal{Z}_{\bs{X}_0}$ conditions are met by which at all times $p_t(\bs{X} \to \infty)$ decays fast enough (e.g. exponentially) so that no probability leak  to infinity occurs, and that a steady ensemble exists. 

In this framework, the average EPR characterizing the CN's dissipation is defined as \cite{schnak}
\begin{multline}
\sigma_t := \sum_\rho \left\langle v_{\rho}(\bs{X})   \ln \frac{v_{+\rho}(\bs{X}) p_t(\bs{X})}{v_{-\rho}(\bs{X} + \nabla_{\!\rho}) p_t(\bs{X}+ \nabla_{\!\rho})}  \right\rangle_t \geq 0  \label{eq:EPRstoc}
\end{multline}
It can easily be proven that the EPR is non-negative, embodying the second law of thermodynamics.  The logarithmic term measures the thermodynamic cost of reaction $\rho$ for a given $\bs{X}$, and it quantifies the degree by which detailed balance is broken.

\subsection{Deterministic EPR}

\label{detEPR}
The corresponding deterministic model is obtained by neglecting correlations and higher cumulants, i.e. by replacing \tc{ $\langle \bs{X}^{\bs{\nu}_\rho} \rangle_t \to (\Omega\bs{x})_t^{\bs{\nu}_\rho}$, where $\Omega$ is a large volume parameter that \aw{makes} $\bs{x}$ a continuous variable \aw{with the interpretaton of a concentration}; in the following we will set $\Omega =1$ for notational clarity and only resume proper scalings when studying the model systems in Sec.\,\ref{numerical}.} Also, in the large volume limit the approximation $v_\rho(\bs{x}) \approx k_{\rho} \bs{x}^{\bs{\nu}_\rho}$ is made. Then Eq.\,(\ref{eq:mean}) yields the rate equation \cite{ross}
\bea
\frac{d\bs{x}_t}{dt}= \sum_{\rho} \bs{\nabla}_{\!\rho} v_\rho(\bs{x}_t) \label{eq:det}
\eea
Again, we are interested in steady behavior, when the right-hand side vanishes. Importantly, while the Chemical Master Equation admits one unique steady ensemble, the corresponding deterministic dynamics might admit none or several locally stable fixed points $\bs{x}_{\infty}$ and more complicated phenomenology such as limit cycles and fractal attractors \cite{ross}. Deterministic multistability corresponds to the steady ensemble being multimodal. Notice that $\bs{x}$ cannot be interpreted as a mean, as for bistable systems the mean might be far from both stable fixed points. Rather, in a scaling limit with the system size, random jump processes can be shown to typically behave deterministically, as rigorously detailed in Ref.\,\cite{andersonbook}.

In this setting, the deterministic EPR is defined as  \cite{beard2}
\bea
\bar{\sigma}_t := \sum_{\rho} v_{\rho}(\bs{x}_t)  \ln \frac{v_{+\rho}(\bs{x}_t) }{v_{-\rho}(\bs{x}_t)} \geq 0. \label{eq:EPRdet}
\eea
The connection to free-energy differences and other thermodynamic potentials in a nonequilibrium setting is detailed in Ref.\,\cite{PolEspo14}.

\begin{figure}
\centering
\includegraphics[width=1.0\columnwidth]{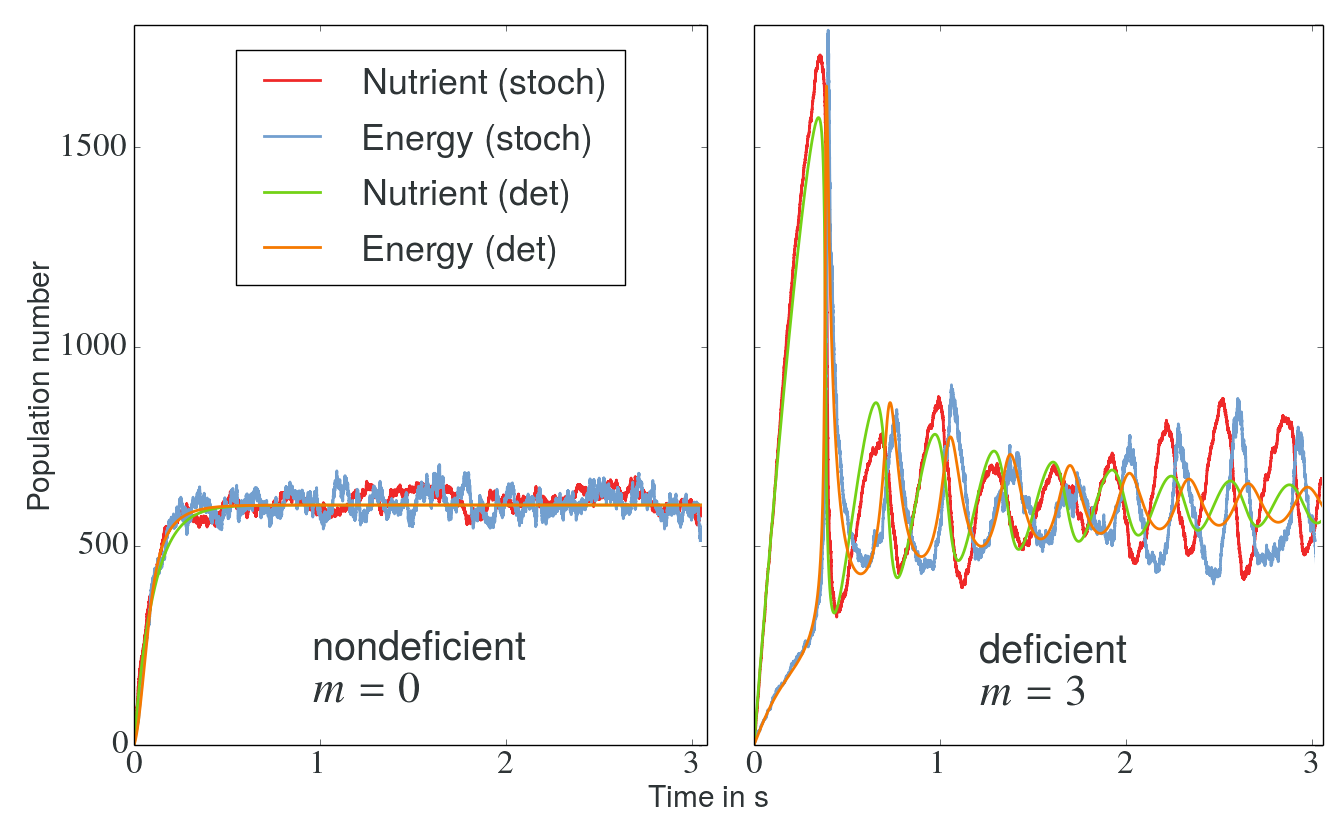}
\caption{ \label{timecourse} We consider a class of toy models for metabolism $\emptyset \leftrightharpoons \mathrm{N}, \mathrm{N} + m\mathrm{E} \leftrightharpoons (m+2)\mathrm{E}, 2\mathrm{E} \leftrightharpoons \emptyset$, for varying $m$. In this figure we compare stochastic and deterministic time evolution of nutrient and energy molecules in the model corresponding to $m=0$, that has deficiency $\delta = 0$ (on the left), and in model corresponding to $m=3$, with deficiency $\delta = 1$ (on the right). The reactor is initially empty; rates are scaled according to the volume-parameter $\Omega = 10^{-21} N_A \approx 602$ (see main text), which is the number of molecules at the fixed point, for both models and for both species. In the zero-deficiency case, stochastic dynamics only adds structure-less noise to the deterministic behavior. Instead, in the deficient case, while the deterministic system has damped oscillations towards the fixed point, oscillations are sustained in the corresponding stochastic dynamics, yielding a structural deviation between the two.}
\end{figure}

\section{Results}

\subsection{Theoretical}

First, we re-work the  above expressions for the deterministic and stochastic EPRs to make them closer one to another. Introducing the {\it thermodynamic forces}
\bea
G_\rho := \ln \frac{k_{+\rho}}{k_{-\rho}},
\eea
that measure the kinetic imbalance of reactions, with a few manipulations we can bring the deterministic EPR to
\bea
\bar{\sigma}_t = \sum_{\rho}  v_\rho(\bs{x}_t)   G_\rho -  \ln \bs{x}_t  \cdot \frac{d\bs{x}_t}{dt} \label{eq:EPRdet2}
\eea
As regards its stochastic counterpart, plugging the mass-action rates, Eq.\,(\ref{eq:LMA}),  into Eq.\,(\ref{eq:EPRstoc}) we obtain 
\bea
\sigma_t =  \sum_{\rho} \langle v_{\rho}\rangle_t  G_\rho - \sum_{\bs{X}} \ln [p_t(\bs{X}) \bs{X}!] \,Lp_t(\bs{X}).
\label{eq:CMEEPR}
\eea
This is the first main result in our paper. Its most remarkable feature is that in the first term, related to the entropy flow to the environment \cite{seifert}, only the ``macroscopic''  average reaction velocity appears, and that ``microscopic'' dependencies on $\bs{X}$ are within the second term, which is related to the system's entropy change. At the trajectory level, this grants the validity of so-called Fluctuation Theorems \cite{PolEspo14b}, hence $\sigma_t$ is a proper notion of EPR. It is important, and {\it a priori} not obvious that the thermodynamic force $G_\rho$ is the same in the stochastic and in the deterministic settings.

Second, we define the correlation EPR as  $\delta\sigma_t :=  \sigma_t - \bar{\sigma}_t$ and notice that, in the steady regime, it can be expressed as a weighted difference between the average and the deterministic reaction velocity, as was anticipated in Eq.\,(\ref{eq:main}). Explicitly, we obtain a formula for the steady correlation EPR as a weighted sum of population moments:
\bea
\delta \sigma_\infty & = & \sum_\rho \Big[ \langle v_\rho \rangle_\infty - v_\rho(\bs{x}_\infty) \Big] G_\rho \\
& = &  \sum_\rho G_\rho k_\rho \Big( \langle \bs{X} \ldots (\bs{X} - \bs{\nu}_\rho + 1 ) \rangle_\infty -\bs{x}_\infty^{\bs{\nu}_\rho}\Big).
\eea
The latter expression might pave the way for approximate estimations of the correlation EPR based on Van Kampen's system size expansion, moment-closure techniques  or other diffusion approximations, provided due care is paid to the fact that such approximations often fail to reproduce the stochastic thermodynamics out of equilibrium \cite{jordan} or even the distibution moments \cite{schnoerr}.

Third, we evaluate the stochastic EPR when the system is in a product-form Poisson-like ensemble\footnote{Notice that, because the range of summation is the lattice orthant $\mathcal{Z}_{\bs{X}_0}$ and not $\mathbb{Z}^{|X|}$, $|X|$ being the number of species, a ``product-form Poisson-like'' distribution is Poissonian in form but not in fact.} with a generic time-dependent parameter $\bs{y}_t$,
\bea
\mathrm{Pois}_{\bs{y}_t}(\bs{X}) = \frac{1}{Z_{\bs{X}_0}} \frac{{\bs{y}_t}^{\bs{X}}}{\bs{X}!},  \label{eq:pois}
\eea
with $Z_{\bs{X}_0}$ the normalization factor over $\mathcal{Z}_{\bs{X}_0}$. 
In this case it can be shown with few manipulations (see Appendix \ref{app3} for a step-by-step derivation) that $\langle v_{\rho}\rangle_{\mathrm{Pois}_{\bs{y}_t}} =  v_{\rho}(\bs{y}_t)$, and consequently
\bea
\sigma_{\mathrm{Pois}_{\bs{y}_t}}  =  \sum_\rho v_{\rho}(\bs{y}_t) G_\rho - \ln \bs{y}_t  \cdot \sum_\rho  \nabla_\rho v_{\rho}(\bs{y}_t). \label{eq:above}
\eea

Notice that this expression coincides with the deterministic EPR at $t \to \infty$ if the Chemical Master Equation admits a steady product-form Poissonian with parameter $\bs{y}_\infty$  being a deterministic fixed point, and at all times if the system admits a product-form Poissonian with time-dependent parameter solving the deterministic rate equations.

Fourth, we investigate under which conditions such hypothesis are met. The ACK theorem \cite{Ander10} entails that, under our reversibility assumption, if the network has null deficiency, then the Chemical Master Equation admits a product-form Poissonian with parameter $\bs{x}_\infty$ being the fixed point of the corresponding deterministic dynamics, which by Feinberg's results \cite{feinberg} for $\delta = 0$ is unique and locally stable. Hence the steady correlation EPR  vanishes for zero-deficiency networks. For sake of reference we sketch a proof of the theorem in Appendix \ref{app2}. Furthermore, it is known that in linear networks where no more than one molecule is consumed or produced at a time (i.e. $\sum_i \nu_{\rho,i} =0,1$), provided the system is prepared in a product-form Poissonian, it maintains such form at all times, with its parameter subjected to the corresponding rate equations \cite{heuett}. Hence for linear CNs prepared in a product-form Poissonian ensemble, the correlation EPR vanishes at all times. These results thus generalize those by Mou et al. \cite{Mou84}, who observed that the correlation EPR vanishes at steady states in linear networks.

\subsection{Numerical}
\label{numerical}

We will now illustrate \tc{the consequences of the ACK theorem and} our findings with the aid of the above class of toy models. In fact we will further simplify the scenario by eliminating the waste $W$, which does not play any substantial kinetic role. Details on the simulation methods can be found in Appendix \ref{app1}.

Let $\Omega$ be a scaling parameter regulating the system's size and let $x = N/\Omega$ be the concentration of N and $y= E/\Omega$ that of E. A convenient choice of parameters is $k_\rho =K_\pm   \Omega^{1-\sum_i \nu_{i\rho}}$, where $K_\pm$ are independent of the reaction, in their respective units (which depend on $\rho$). Then for given $\Omega$ all models turn out to have the same fixed point concentrations and steady EPR, making them easily comparable. Concentrations obey the system of rate equations
\begin{equation}
\begin{aligned}
\dot{x} & = K_+ - K_-  x -  K_+ x y^m + K_- y^{n+m} \\
\dot{y} & = n \left(K_+ x y^m - K_- y^{n+m} + K_- - K_+y^n\right).
\end{aligned}
\end{equation}
A fixed point is found at $x_\infty = y_\infty = 1$, for all values of $m,n$. Its stability depends on $m,n,K_+,K_-$. The deterministic EPR at the fixed point is given by
\bea
\bar{\sigma}_\infty = 3\Omega (K_+ - K_-) \ln \frac{K_+}{K_-} \label{eq:detEPex}
\eea
(notice that parameter $\Omega$ cancels within the logarithms, so that the EPR is extensive) and again it is independent of $m,n$.

We will consider the cases $n=2$, for values $m=0,1,2,3$, $m=0$ being the zero-deficiency case, all others having $\delta = 1$. We take $K_+ = 10$, $K_- =1$, which signifies that the system is very far from a detailed balanced thermodynamic equilibrium. We start from an empty reactor,  $x_0=y_0 =0$. For these values the above fixed point is stable for all $m < 4 $. For $m=0$ the dynamics converges uniformly to the fixed point, as shown in the left-hand side of Fig.\,\ref{timecourse}. A more interesting behavior appears for higher $m$: for $m=3$ the deterministic system displays damped oscillations towards the fixed point (as shown by the innermost smoother lines in the left-hand side of Fig.\,\ref{timecourse}). Indeed, for $m=4$ the fixed point becomes unstable and the system displays steady oscillations.

\begin{figure}
\centering
\includegraphics[width=1.0\columnwidth]{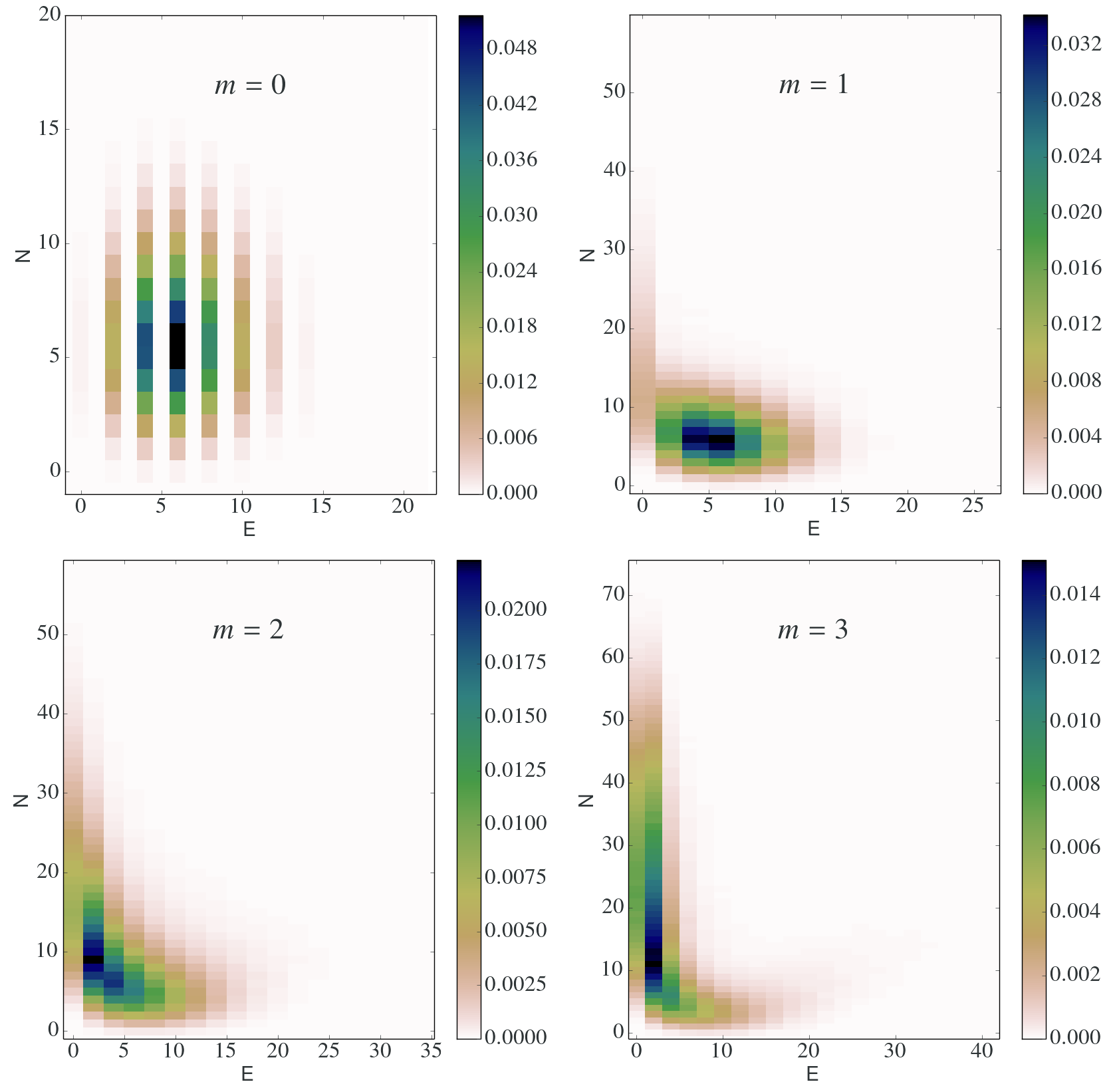}
\caption{ \label{bihisto} The ACK theorem states that, if a CN has zero deficiency, then given an initial state (in our case, $N=E=0$), the steady ensemble of the Chemical Master Equation has product form. Here we display color-plots of the histograms of the steady distribution of nutrient and energy molecules, for our toy models with $m=0,1,2,3$, and rates scaled down by the volume-parameter  $\Omega = 10^{-23} N_A = 6.02$, giving a low number of molecules at the steady state. Zebra lines (present, but not displayed for $m>0$ for sake of better visualization) indicate that the stochastic dynamics preserves the parity of the energy molecules, which are produced in pairs. Owing to the outer smudge, the deficient models $m=1,2,3$ have a non-product form distribution. The product-form distribution of the zero-deficiency case $m=0$ is shown in more detail in Fig.\,\ref{histo}.}
\end{figure}

\begin{figure}
\centering
\includegraphics[width=1.0\columnwidth]{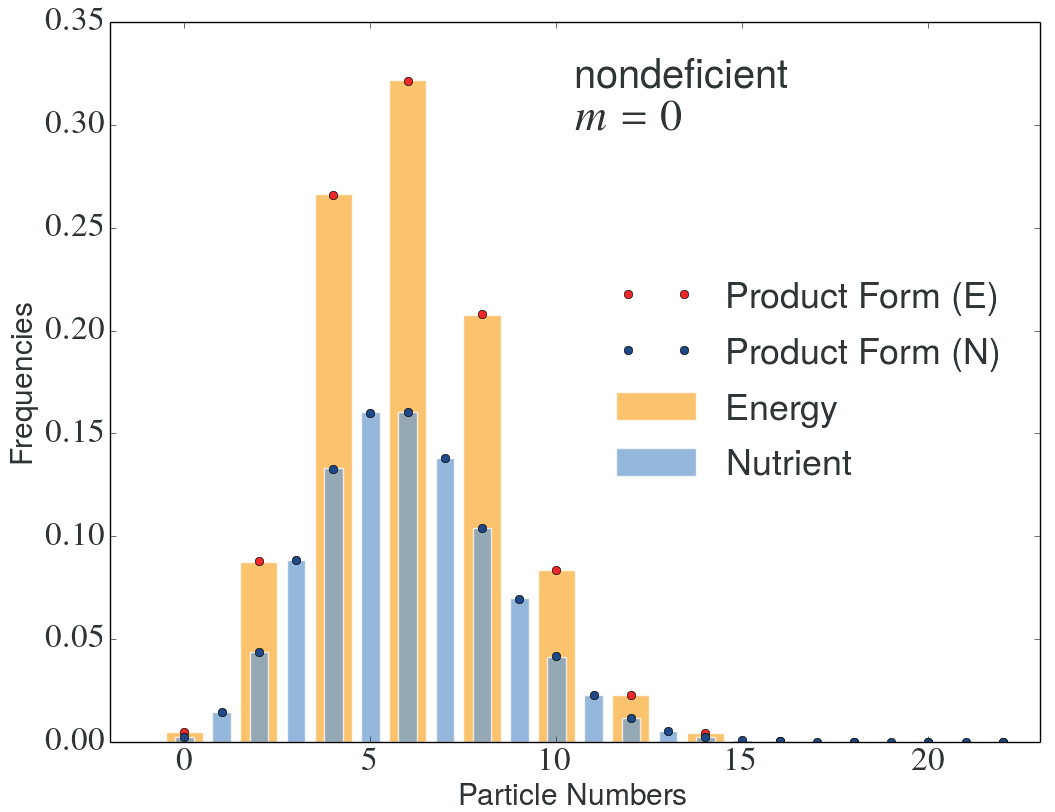}
\caption{ \label{histo} The nonlinear CN $\emptyset \leftrightharpoons \mathrm{N} \leftrightharpoons 2\mathrm{E} \leftrightharpoons \emptyset$ (corresponding to $m=0$) has zero deficiency. Hence, by the ACK theorem its corresponding Chemical Master Equation affords a product-form steady ensemble, and the marginals for the number of nutrients and of energy molecules also have Poisson-like distributions. We plot histograms for the populations of nutrient and energy molecules generated by stochastic simulations via Gillespie's algorithm, with rates scaled by a volume parameter $\Omega = 6.02$, showing perfect agreement with the predictions of the ACK theorem.}
\end{figure}

As regards the stochastic setting, so far our framework was that of ensemble thermodynamics, describing a large sample of processes at a given time. From now on we consider one given process in a large time. Indeed, Stochastic Thermodynamics has two complementary formulations: one along ensembles, and one along individual processes  \cite{broeck}. The two frameworks are compatible, since the ergodic principle ensures that long-time averages almost surely (a.s.) equal ensemble averages at the steady state. In particular it can be proven that for the reaction velocity
\bea
\langle v_\rho \rangle_\infty = \lim_{t \to \infty} \frac{1}{t} \#_t (\rho), \quad a.s. 
\eea
where $\#_t (\rho) $ is the number of times reaction $\rho$ has been performed along the stochastic trajectory up to time $t$. Similarly, a histogram for the steady ensemble $p_\infty(N,E)$ can be obtained by calculating the average time spent by the trajectory at state $N,E$. Let us then illustrate the ACK theorem. In Fig.\,\ref{bihisto} we provide color-plots for $p_\infty(N,E)$. For $m=0$, the color plot renders the distribution's product-form. Zebra-lines are due to the fact that energy tokens are produced in pairs, hence starting from $x_0=y_0 =0$ only even numbers of energy molecules can be populated. The same zebra-structure occurs for higher $m>0$, but for sake of better visualization we drew pixels twice the width, covering the whole area. The smudge in the color plots in Fig.\,\ref{bihisto} for $m> 0$ reveals that the steady ensemble does not have product form. Instead, in the zero-deficiency case, Fig.\,\ref{histo} compares the histograms of the marginals for the energy and the nutrient, showing that they perfectly agree with the prediction from the product-form Poissonian.

In Fig.\,\ref{thermo} we plot the average stochastic EPR as a function of volume $\Omega$. The perfect overlap between the deterministic EPR (upper line) and the dots corresponding to the $m=0$ case confirms our result that for deficiency-zero systems the correlation EPR vanishes. For $m>0$ this particular class of models has negative correlation EPR. The plots of the relative error in the inset show that the effect vanishes at large system sizes where fluctuations become negligible.

\begin{figure}
\centering
\includegraphics[width=1.0\columnwidth]{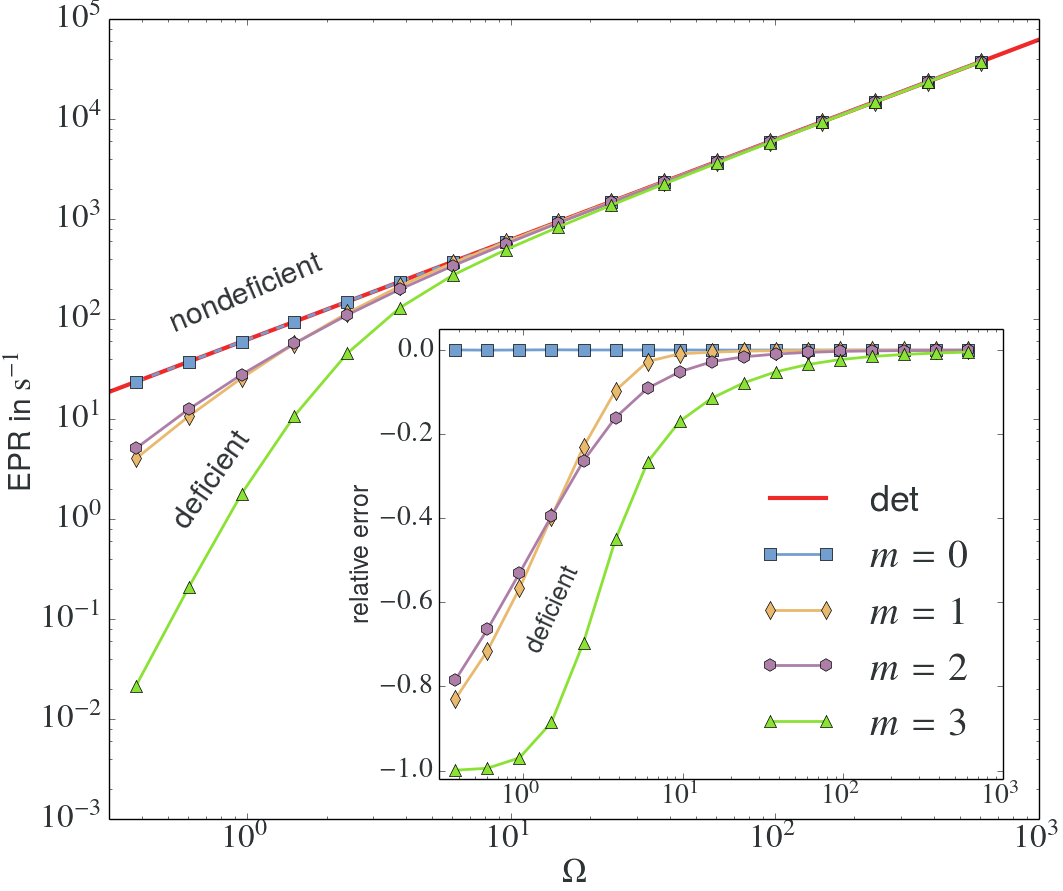}
\caption{ \label{thermo} The main result of our paper is that dissipation (EPR) in stochasic chemical dynamics only coincides with the deterministic EPR when the CN has zero deficiency, and that already in simple systems intrinsic noise affects dissipation. In the main frame we plot in log-log scale the stochastic EPR for our toy models, for all values $m=0,1,2,3$, as a function of the volume-parameter $\Omega$ that sets the average number of molecules present in the reactor at the steady ensemble. The upper straight line represents the deterministic value, Eq.\,(\ref{eq:detEPex}). The dots on top of it are the values of the corresponding stochastic zero-deficiency system, $m=0$. Models $m\geq 1$ with deficiency $\delta =1$ have lower EPR than the deterministic model. An explanation for this is in Fig.\,\ref{lattice}. In the inset, we show that the relative error between stochastic and deterministic values decreases with volume.}
\end{figure}

\begin{figure}
\centering
\includegraphics[width=1.0\columnwidth]{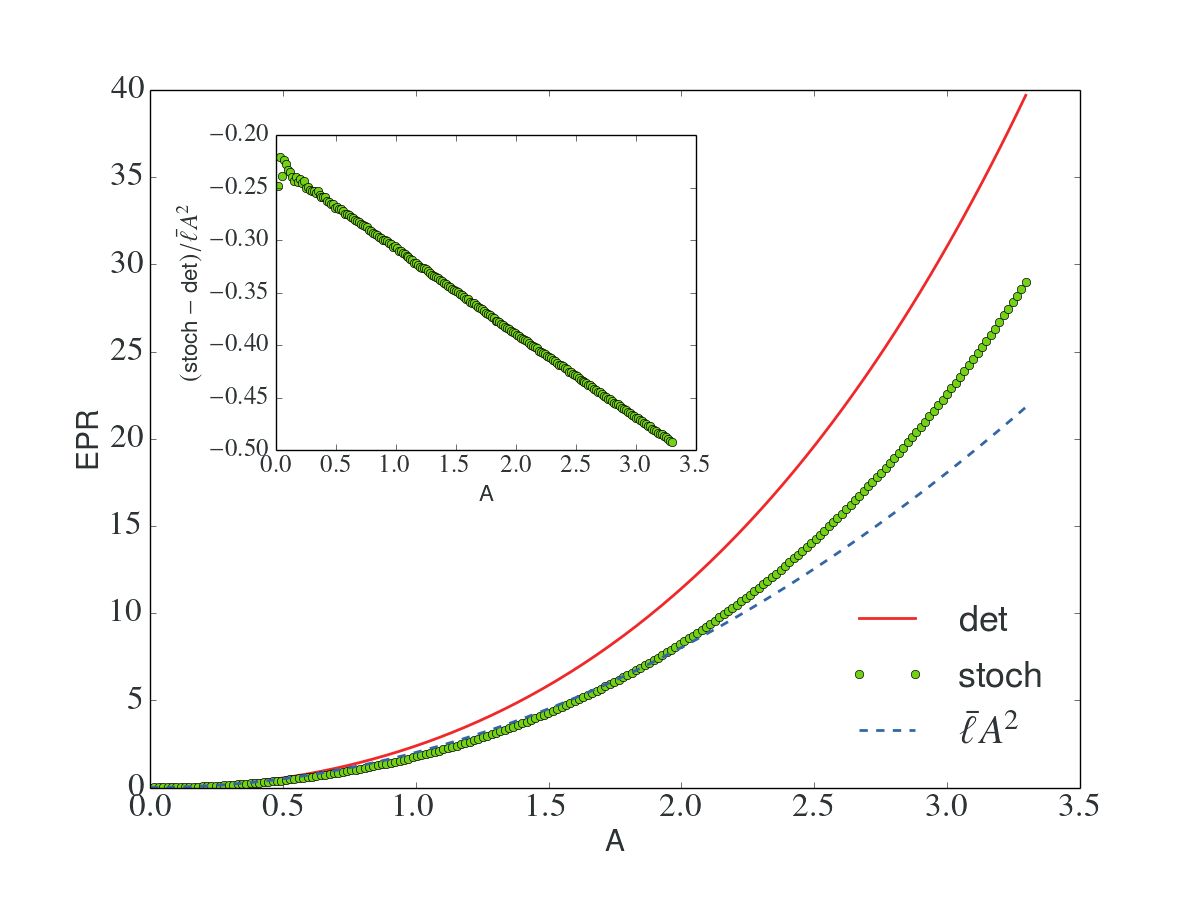}
\caption{\tc{The affinity $A = 3 \log K_+/K_-$ determines the distance from thermodynamic equilibrium (detailed balance). In this figure we show the dependency of the deterministic and the stochastic EPRs with respect to the affinity, for $m = 3$ and $\Omega = 6.02$, at fixed $K_- =1$ and variable $K_+$. The \aw{dashed} curve is the linear-regime approximation of the deterministic EPR, where the current is approximately linear in the affinity and the EPR is approximated by a quadratic. Clearly the EPRs approach zero for vanishing affinity (no dissipation). The inset shows the error between stochastic and deterministic EPR, relative to the linear approximation. The relative error increases with the affinity and, remarkably, it does not tend to vanish for $A \to 0$. This implies that for nonvanishing deficiency, the Onsager coefficients of the deterministic and stochastic systems differ.}  \label{affinity}}
\end{figure}

{\color{black} Finally, another interesting aspect to inquire is the dependency of the correlation EPR on the affinity $A = 3 \log K_+/K_-$, which determines the distance from detailed balance, i.e. from thermodynamic equilibrium. In particular, we are interested in the so-called {\it linear regime} where the affinity is small and stationary currents are approximately linear in the affinity. Then
\bea
\delta \sigma_{\infty} = (\ell - \bar{\ell}) A^2
\eea
with the deterministic {\it linear response coefficient} $\bar{\ell} = \Omega/3$. The inset in Fig.\,\ref{affinity} shows that in a model with nonvanishing deficiency, in the linear regime the correlation EPR, relative to the deterministic linear regime approximation, does not vanish in the limit $A\to 0$, which implies that the stochastic linear response coefficient $\ell$ differs from the deterministic one.}

\tc{Our result proves that having $\delta =0$ is a sufficient condition for a vanishing correlation EPR. A preliminary question is then whether it is also necessary. The answer is trivially negative. In fact, if rates are such that detailed balance holds, then both the stochastic, the deterministic, and hence the correlation EPRs vanish. More generally, for the ACK theorem to hold it is sufficient that the more general condition of {\it complex balance} holds: even if deficiency is greater than zero, rates can conjure in such a way that currents look ``as if'' the system had null deficiency. Furthermore, by the theory of Schnakenberg\cite{schnak} it can be shown that the correlation EPR can be decomposed in fundamental cycles $\delta \sigma(\infty) = \sum_\alpha \left[ \langle J_{\alpha} \rangle_\infty - J_{\alpha}(\bs{x}_\infty) \right] A_\alpha$, with index $\alpha$ spanning a basis of the null space of the stoichiometric matrix, $A_\alpha$ a {\it cycle affinity} and $J_\alpha$ a {\it cycle current}. Cycle affinities are invariant under a wide range of transformations of the rate constants which affect the cycle currents; hence even for non-complex balanced rates it might be feasible to tune the rates in such a way that several cycle contributions all cancel each other.}

\tc{The above argument rests on the fact that rate constants might be fine-tuned. The question becomes more interesting if properly reformulated. For systems with nonvanishing deficiency, complex-balanced rates are a set of measure zero in the space of possible rates. So, is the condition $\delta = 0$ necessary for a vanishing correlation EPR, for all possible values of rates? Very special systems with nonvashing deficiency which still have Poissonian steady states have been found\cite{cappelletti}. An example is the chemical network $\mathrm{X} + \mathrm{Y} \rightleftharpoons 2\mathrm{X} +  \mathrm{Y} $, $\mathrm{X} \rightleftharpoons 2\mathrm{X}$. In this case, the number of molecules of $\mathrm{Y}$ is constant and determines the stoichiometric compatibility class where the dynamics is restricted. The deficiency is $\delta = 1$, still the steady ensemble is a product-form Poissonian with parameter given by the solution of the deterministic equations of motion, and the correlation EPR can be easily shown to vanish. To take this class of cases into the description, Cappelletti and Wiuf have introduced the concept of ``stochastically complex-balanced'' chemical reaction networks. The analysis of whether correlation EPR vanishes for all values of the rates if and only if the network is stochastically complex-balanced goes beyond the scope of the present paper.}

\section{Discussion and conclusions}

While it could have been expected that fluctuations would increase dissipation, our simple model displays the opposite behavior. This can be explained as follows. Notice that for $m=3$ in Fig.\,\ref{timecourse} the stochastic dynamics has amplified oscillations, such as those characterized in Ref.\,\cite{mckane}, where a purely stochastic mechanism for biochemical oscillations was proposed.  Such oscillations are forcedly stabilized in the deterministic setting. Hence the stochastic model is more flexible and capable of exploring modes that the deterministic system abandons. Lower EPR then occurs when such modes are entropically convenient. A way to characterize these modes is by a switching mechanism of chemical pathways. Fig.\,\ref{lattice} details that in deficient networks, at low molecule numbers certain reactions can be effectively shut off because of the temporary absence of a sufficient number of reactants. This phenomenon eventually reshapes the structure of the irreversible closed reaction pathways that the system can locally perform. In our particular model, for low molecule numbers reaction $2$ is inhibited, and the other two reactions alone do not contribute to dissipation. Instead, in the CN with $\delta =0$ the dissipative cycle can be performed at any particle number. 
 
\tc{The above example might then lead to hypothesize that the correlation EPR could be non-positive \aw{in general}. This is not the case though. A counterexample can be found in the literature. The Schl\"ogl model $\emptyset \rightleftharpoons X$, $2X \rightleftharpoons 3X$ has deficiency $\delta =1$, and its most important feature is that for certain critical values of the parameters it displays a bifurcation. Gaspard compared stochastic and deterministic EPRs for this model \cite{gaspard}, and as can be observed from Fig.\,2 in Ref.\cite{gaspard}, close to the critical point the stochastic EPR is larger than the deterministic one, while in the \aw{bistable} region it interpolates between the two possible values that the deterministic EPR takes at each of the two stable fixed points.}

Despite the fact that our toy model is oversimplified, the mechanisms we observed might carry out to more realistic networks. At the level of gene expression, it is known that intrinsic noise is a crucial factor in phenotypic variation within isogenic populations \cite{elowitz}.  One step below, while in cells metabolites might be large in number, gene-expressed regulatory molecules might be very few\footnote{In {\it E. coli}, the lowest-concentration metabolite, nucleoside adenosine, is present in $\sim 10^2$ copies, but over 80\% of the variety proteins is much lower in copy numbers \cite{guptasarma}}, allowing the switching mechanisms that we described above. In metabolism, the action of enzymes typically adds a level of complexity. In fact, most (if not all) of the reactions in biochemical CNs are not elementary, hence their connectivity and kinetic rules have to be determined {\it a posteriori} by advanced experimental methods (see \cite{vance} for a systematic review). Nevertheless, in our models the inbuilt deficient cycle could be seen as the core structure of any metabolic model. The network should be enriched by resolving individual metabolites within nutrients and waste, adding intermediate reactants such as cofactors and enzymes, resolving the environment and outer thermodynamic cycles, separating time-scales and resorting to effective rate laws when applicable. As a proof of concept, all these operations will in general maintain the core cycle and hence the deficient character of the network, hence it can be argued that, because of its autocatalytic character, metabolism is deficient.

To conclude, we emphasize that understanding thermodynamic constraints on the regulation of metabolic networks is a crucial problem in CN reconstruction \cite{qianMN,TMFA}. In this work we displayed a close connection between the topological notion of deficiency of a CN and nonequilibrium thermodynamics, proving that at steady states only in zero-deficiency CNs the EPR evaluated by the mean-field deterministic theory coincides with that of the corresponding stochastic model, accounting for stochastic variability in molecules' number at low concentrations. For deficient CNs a nonvanishing correlation EPR quantifies the disagreement between deterministic and stochastic modeling, and at low molecule numbers this disagreement can be understood in terms of a switching mechanisms of reaction pathways. A more detailed study of the conditions for positive vs. negative correlation EPR is demanded to future inquiry. Immediate perspectives also include the study of non-well-stirred mixtures, where reaction-diffusion processes allow for pattern formation, and of systems with separation of time scales and effective enzymatic reactions. On the computational side, the more demanding stochastic techniques can be blended with deterministic algorithms to provide efficient tools for the systematic computation of the entropic balance of a CN, e.g. in software like COPASI \cite{copasi}. More work has to be done to delineate future application of deficiency theory and stochastic thermodynamics to realistic metabolic networks.

\begin{figure}
\centering
\includegraphics[width=.8\columnwidth]{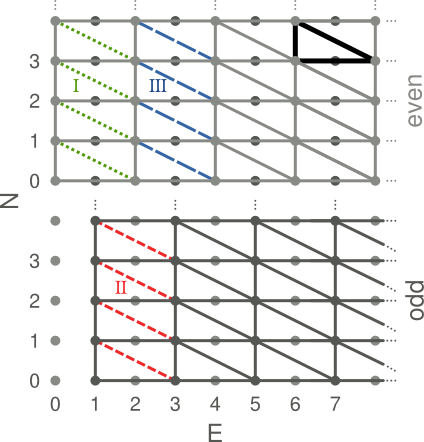}
\caption{\label{lattice} Chemical stochastic kinetics occurs on lattice orthants, called  stoichiometric compatibility classes (SCC). For our class of models, given an initial state, random jumps preserve the parity of the energy molecules (even or odd), hence there are two distinct SCCs. In the zero-deficiency case, $m=0$, all of the drawn transitions are possible, and both SCCs can be obtained by repeatedly copy-pasting a motif corresponding to the full CN, marked bold in the figure,  through the whole lattice orthant. That is, locally each SCC looks like the full CN. Only cycling trajectories that carry a thermodynamic affinity contribute to the steady stochastic EPR \cite{schnak,PolEspo14b}. Hence, for $m=0$, even for very low molecule numbers it is always possible to perform the entropy-producing cycle. On the other hand, the structure of the SCCs for deficient networks is: for $m=1$ dotted transitions type I are not feasible (since at least one energy token is needed to perform reaction $\rho = + 2$ and three energy tokens are needed to perform $\rho = -2$), for $m=2$ dotted transitions type I and II are switched off, and for $m=3$ transitions type I, II and III are shut. Hence for low-enough molecule numbers, the stochastic trajectory explores a portion of the SCC where there is no possibility of producing entropy along an irreversible cycle (cycles consisting only of reactions $\rho =\pm 1,\pm 2$ don't dissipate). This explains the lower stochastic EPR observed for $m=1,2,3$ in Fig.\,\ref{thermo}.}
\end{figure}

\paragraph*{Aknowledgments.} The research was supported by the National Research Fund Luxembourg in the frame of project FNR/A11/02, of the AFR Postdoc Grant 5856127 and of the AFR Ph.D. Grant 7865466.

\appendix

\section{\label{app3} Explicit derivation of Eq.\,(\ref{eq:above})}

From Eq.\,(\ref{eq:EPRstoc}), plugging into the rates Eq.\,(\ref{eq:LMA}) and the Poisson-form distribution Eq.\,(\ref{eq:pois}) we obtain
\bea
\sigma_{\mathrm{Pois}_{\bs{y}_t}}  & = & \sum_\rho \sum_{\bs{X}} \mathrm{Pois}_{\bs{y}_t}(\bs{X}) v_{\rho}(\bs{X})\nonumber  \\
& & \qquad   \ln \frac{v_{+\rho}(\bs{X})  \mathrm{Pois}_{\bs{y}_t}(\bs{X})}{v_{-\rho}(\bs{X} + \nabla_{\!\rho})
 \mathrm{Pois}_{\bs{y}_t}(\bs{X}+ \nabla_{\!\rho})} \nonumber \\
& = & \frac{1}{Z_{\bs{X}_0}}  \sum_\rho \sum_{\bs{X}}   k_{\rho}  \frac{{\bs{y}_t}^{\bs{X}}}{(\bs{X} - \bs{\nu}_{\rho})!} \ln \frac{ k_{\rho} \frac{{\bs{y}_t}^{\bs{X}}}{(\bs{X} - \bs{\nu}_{\rho})!} }
{k_{-\rho} \frac{{\bs{y}_t}^{\bs{X}+ \nabla_{\!\rho}}}{(\bs{X} + \nabla_{\!\rho} - \bs{\nu}_{-\rho})!}} \nonumber \\
& = & \frac{1}{Z_{\bs{X}_0}}  \sum_\rho \sum_{\bs{X}}   k_{\rho}  \frac{{\bs{y}_t}^{\bs{X}}}{(\bs{X} - \bs{\nu}_{\rho})!} \left( \ln \frac{ k_{\rho}}{k_{-\rho}} - \nabla_{\!\rho} \cdot \ln {\bs{y}_t} \right) \nonumber  
\eea
We now shift the summation over $\bs{X}$ to obtain
\bea
\sigma_{\mathrm{Pois}_{\bs{y}_t}}
& = & \frac{1}{Z_{\bs{X}_0}} \sum_{\bs{X}}
 \frac{{\bs{y}_t}^{\bs{X}}}{\bs{X}!} \sum_\rho   k_{\rho} 
{\bs{y}_t}^{\bs{\nu}_{\rho}}\left( \ln \frac{ k_{\rho}}{k_{-\rho}} - \nabla_{\!\rho} \cdot \ln {\bs{y}_t} \right) \nonumber \\
& = & \sum_\rho   k_{\rho} 
{\bs{y}_t}^{\bs{\nu}_{\rho}}\left( \ln \frac{ k_{\rho}}{k_{-\rho}} - \nabla_{\!\rho} \cdot \ln {\bs{y}_t} \right) \nonumber
\eea
which is the desired result, Eq.\,(\ref{eq:above}).

\section{\label{app2} Sketch of derivation of the deficiency-zero theorem}

One of the corollaries that incarnate the Anderson-Craciun-Kurtz theorem \cite{Ander10}  states that if a (weakly) reversible CN has deficiency zero, then on each stoichiometric compatibility classes the Chemical Master Equation admits a product-form Poisson-like steady distribution with parameter given by the unique fixed point of the corresponding rate equations. For sake of completeness, we provide the sketch of a derivation based on the graph-theoretical perspective that was briefly introduced in the main text. For another derivation based on quantum techniques, see \cite{baez}.

Plugging the product-form Eq.\,(\ref{eq:pois}) with parameter given by the fixed point $\bs{x}_\infty$ into the generator Eq.\,(\ref{eq:CME}), and using rates Eq.\,(\ref{eq:LMA}) one obtains

\bea
& & {L}\mathrm{Pois}_{\bs{x}_{\infty}}(\bs{X}) = \nonumber \\
& = &\frac{1}{Z_{\bs{X}_0}} \sum_{\rho} \Big[
k_{-\rho} \frac{{\bs{x}_{\infty}}^{\bs{X}+\bs{\nabla}_{\!\rho}}}{(\bs{X} +  \bs{\nabla}_{\!\rho} - \bs{\nu}_{-\rho})!} -  k_{+\rho} \frac{{\bs{x}_{\infty}}^{\bs{X}}}{(\bs{X} - \bs{\nu}_{\rho})!} 
\Big] \nonumber \\ 
& = & \frac{2}{Z_{\bs{X}_0}} \sum_{\rho > 0} \frac{{\bs{x}_{\infty}}^{\bs{X}-\bs{\nu}_\rho}}{(\bs{X} - \bs{\nu}_{\rho})!}  \Big[
v_{-\rho}  (\bs{x}_{\infty}) -  v_{+\rho}(\bs{x}_{\infty})
\Big]
\eea
where we used $ \bs{\nabla}_{\!\rho} = \bs{\nu}_{-\rho} - \bs{\nu}_{\rho}$, and antisymmetrized. We now observe that the sum over reaction vectors $\rho >0$ can be commuted with a sum over complexes $\mathrm{Y}_i$, followed by a sum over all reactions $\rho$ that have $\mathrm{Y}_i$ as a source complex. The latter information is stored into the incidence matrix $\partial$ of the graph of complexes. Noticing that $\nu_\rho$ only depends on the complex of reactants ahead of $\rho$, we can write
\begin{multline}
{L}\mathrm{Pois}_{\bs{x}_{\infty}}(\bs{X}) =  \frac{2}{Z_{\bs{X}_0}} \sum_{i} \frac{{\bs{x}_{\infty}}^{\bs{X}-\bs{\nu}_i}}{(\bs{X} - \bs{\nu}_i)!}  \\ \sum_\rho \partial_{i,\rho} \Big[
v_{-\rho}  (\bs{x}_{\infty}) -  v_{+\rho}(\bs{x}_{\infty})\Big]. \label{eq:comcom}
\end{multline}
After Eq.\,(\ref{eq:det}), the fixed point satisfies
\bea
\sum_{\rho >0} \bs{\nabla}_{\!\rho} \left[ v_{+\rho}(\bs{x}_\infty) - v_{-\rho}(\bs{x}_\infty) \right] = 0
\eea
which implies that $v_{+\rho}(\bs{x}_\infty) - v_{-\rho}(\bs{x}_\infty)$ is a right-null vector of the stoichiometric matrix. But if $\delta =0$, then $v_{+\rho}(\bs{x}_\infty) - v_{-\rho}(\bs{x}_\infty)$ is also a right-null vector of the incidence matrix (see last paragraph in Sec.\,\ref{meta}), hence Eq.\,(\ref{eq:comcom}) vanishes.

\section{\label{app1} Materials and methods}

We employed the CN simulation software COPASI \cite{copasi} to simulate the Chemical Master Equation via Gillespie's algorithm, and the LSODA algorithm implemented in the scientific python stack (SciPy) \cite{scipy} to solve deterministic rate equations. Histograms in Fig.\,\ref{histo} and Fig.\,\ref{bihisto} were sampled from stochastic trajectories for random-time change Markov jump processes spanning over \SI{e5}{\second} with a time resolution of \SI{e-1}{\second}, for a total of \num{e6} binned particle number pairs, while the stochastic time-courses in Fig.\ref{timecourse} employ trajectories of \SI{5}{\second} with a resolution of \SI{e-5}{\second}. Each value for the average stochastic EPR in Fig.\,\ref{thermo} was calculated along single simulations of \SI{e5}{\second}.  Notice that Gillespie's algorithm keeps track of all reaction events, hence the final result for the stochastic average EPR is independent of time resolution. For the deterministic transients we used the same time-span and resolution as for the stochastic ones.  The deterministic EPR was calculated via Eq.\,(\ref{eq:detEPex}) and not from the simulation data. Thus it is only valid at the fixed point.


\begin{thebibliography}{10}

\bibitem{baart} G.J.E.~Baart and D.E.~Martens, {\em Genome-scale metabolic models: reconstruction and analysis}, in  {\em Neisseria meningitidis} (Humana Press, 2012), pp.~107--126.

\bibitem{human} I.~Thiele et al., {\em A community-driven global reconstruction of human metabolism}, Nat. Biotechnol.  {\bf 31}, 419  (2013).


\bibitem{elowitz} M.E.~Elowitz, A.J.~Levine, E.D.~Siggia and P.S.~Swain, {\em Sochastic gene expression in a single cell}, Science {\bf 297}, 1183  (2002).

\bibitem{levine} E.~Levine and T.~Hwa, {\em Stochastic fluctuations in metabolic pathways}, Proc. Natl. Acad. Sci. USA {\bf 104}, 9224  (2007).

\bibitem{tans} D.J.~Kiviet, P.~Nghe, N.~Walker, S.~Boulineau, V.~Sunderlikova,	and S.J.~Tans, {\em Stochasticity of metabolism and growth at the single-cell level}, Nature {\bf 514}, 376  (2014).

\bibitem{paulsson1} I.~Lestas, J.~Paulsson, N.E.~Ross, and G.~Vinnicombe, {\em Noise in gene regulatory networks}, IEEE {\bf 53}, 189 (2008).

\bibitem{broeck} C.~Van den Broeck and M.~Esposito, {\em Ensemble and trajectory thermodynamics: A brief introduction}, Physica A {\bf 418}, 6 (2014). 

\bibitem{ross} J.~Ross, {\em Thermodynamics and Fluctuations far from Equilibrium}, (Springer-Verlag Berlin Heidelberg 2008).

\bibitem{seifert} T.~Schmiedl and U.~Seifert, {\em Stochastic thermodynamics of chemical reaction networks}, J. Chem. Phys. {\bf 126}, 044101 (2007).

\bibitem{PolEspo14}
M.~Polettini and M.~Esposito, {\em Irreversible thermodynamics of open chemical networks I: Emergent cycles and broken conservation laws}, J. Chem. Phys. {\bf 141}, 024117  (2014).

\bibitem{jordan} J.M. Horowitz, {\em Diffusion approximations to the chemical master equation only have a consistent stochastic thermodynamics at chemical equilibrium}, J. Chem. Phys. {\bf 143}, 044111 (2015).

\bibitem{ghosh} A.~Ghosh, {\em Non-equilibrium dynamics of stochastic gene regulation}, J. Biol. Phys. {\bf 41}, 49  (2015).

\bibitem{mehta} P.~Mehta and D.J.~Schwab, {\em Energetic costs of cellular computation},  Proc. Natl. Acad. Sci. USA, {\bf 109}, 17978 (2012).

\bibitem{andrieuxPNAS} D.~Andrieux and P.~Gaspard,  {\em Nonequilibrium generation of information in copolymerization processes}, Proc. Natl. Acad. Sci. USA {\bf 105}, 9516  (2008).

\bibitem{rao} R.~Rao and L.~Peliti, {\em Thermodynamics of accuracy in kinetic proofreading: Dissipation and efficiency trade-offs}, J. Stat. Mech.  P06001 (2015).

\bibitem{gaspard} P.~Gaspard,  {\em Fluctuation theorem for nonequilibrium reactions}, J. Chem. Phys. {\bf 120}, 8898  (2004).

{\color{black}
\bibitem{qian2} H.~Qian, {\em Phosphorylation energy hypothesis: open chemical systems and their biological functions}, Annu. Rev. Phys. Chem. {\bf 58}, 113 (2007).
}

\bibitem{feinberg} M.~Feinberg, {\em Chemical reaction network structure and the stability of complex isothermal reactors-I. The deficiency zero and deficiency one theorems}, Chem. Eng. Sci. {\bf 42}, 2229 (1987).

\bibitem{craciun06} G.~Craciun, Y.~Tang, and M.~Feinberg, {\em Understanding bistability in complex enzyme-driven reaction networks}, Proc. Natl. Acad. Sci. USA {\bf 103}, 8697 (2006).

\bibitem{acgw} D.F.~Anderson, G.~Craciun, M.~Gopalkrishnan, C.~Wiuf, {\em Lyapunov functions, stationary distributions, and non-equilibrium potential for chemical reaction networks}, Bull. Math. Biol., 1 (2015).



\bibitem{tyson} J.J.~Tyson, R.~Albert, A.~Goldbeter, P.~Ruoff, and J.~Sible, {\em Biological switches and clocks}, J. R. Soc. Interface {\bf 5}, S1 (2008).

\bibitem{Ander10} D.F.~Anderson, G.~Craciun, and T.~Kurtz, {\em Product-form stationary distributions for deficiency zero chemical reaction networks}, B. Math. Biol. {\bf 72}, 1947 (2010).

\bibitem{Mou84} C.Y.~Mou, J.-L.~Luo and G.~Nicolis, {\em Stochastic thermodynamics of nonequilibrium steady states in chemical reaction systems}, J. Chem. Phys. {\bf 84}, 7011 (1986).

\bibitem{schaft} A.~van der Schaft, S.~Rao, and B.~Jayawardhana, {\em On the mathematical structure of balanced chemical reaction networks governed by mass action kinetics}, SIAM J. Appl. Math. {\bf 73}, 953 (2013).

\bibitem{schnak} J.~Schnakenberg, {\em Network theory of microscopic and macroscopic behavior of master equation systems}, Rev. Mod. Phys. {\bf 48}, 571  (1976). 

\bibitem{andersonbook} D.F.~Anderson and T.G.~Kurtz, {\em Stochastic Analysis of Biochemical Systems} (Springer, 2015).

{\color{black} 
\bibitem{beard2} H.~Qian and D.A.~Beard, {\em Thermodynamics of stoichiometric biochemical networks in living systems far from equilibrium}, Biophys. Chem. {\bf 114}, 213 (2005).
}

\bibitem{PolEspo14b} M.~Polettini and M.~Esposito, {\em Transient fluctuation theorem for the currents and initial equilibrium ensembles}, J. Stat. Mech. P10033  (2014).

\bibitem{schnoerr} D.~Schnoerr, G.~Sanguinetti and R.~Grima, {\em Validity conditions and stability of moment closure approximations for stochastic chemical kinetics}, J. Chem. Phys. {\bf 141}, 084103 (2014).

\bibitem{heuett} W.J.~Heuett and H.~Qian, {\em Grand canonical Markov model: A stochastic theory for open nonequilibrium biochemical networks}, J. Chem. Phys. {\bf 124}, 044110  (2006).

\bibitem{guptasarma} P.~Guptasarma, {\em Does replication-induced transcription regulate synthesis of the myriad low copy number proteins of Escherichia coli?}, Bioessays {\bf 17}, 987 (1995).

\bibitem{vance} W.~Vance, A.~Arkin and J.~Ross, {\em Determination of causal connectivities of species in reaction networks},  Proc. Natl. Acad. Sci. USA, {\bf 99} (2002), pp.~5816--5821.

\bibitem{qianMN} D. A.~Beard, S.-D.~Liang, and H.~Qian, {\em Energy balance for analysis of complex metabolic networks}, Biophys. J. {\bf 83}, 79 (2002).

\bibitem{TMFA} C.S.~Henry, L.J.~Broadbelt, and V.~Hatzimanikatisy, {\em Thermodynamics-Based Metabolic Flux Analysis}, Biophys. J. {\bf 92}, 1792 (2007).

{\color{black} 
\bibitem{cappelletti} D.~Cappelletti and C.~Wiuf, {\em Product-form Poisson-like distributions and complex balanced reaction systems}, arXiv:1507.02195 (2015).
}

\bibitem{mckane} A.J.~McKane, J.D.~Nagy, and M.O.~Stefanini, {\em Amplified biochemical oscillations in cellular systems}, J. Stat. Phys. {\bf 128}, 165 (2007). 

\bibitem{copasi} S.~Hoops, S.~Sahle, R.~Gauges, C.~Lee, J.~Pahle, N.~Simus, M.~Singhal, L.~Xu, P.~Mendes and U.~Kummer, {\em COPASI: a COmplex PAthway SImulator}, Bioinformatics {\bf 22}, 3067 (2006).

\bibitem{scipy} K.J.~Millman, M.~Aivazis, {\em Python for Scientists and Engineers}, Computing in Science \& Engineering {\bf 13}, 9 (2011).

\bibitem{baez} J.C.~Baez and B. Fong, {\em Quantum techniques for studying equilibrium in reaction networks}, Journal of Complex Networks, {\bf 3}, 22 (2015). 





\end{thebibliography}
\end{document}